# AI and Tempo Estimation: A Review

Geoff Luck[1]

[1] *Centre of Excellence in Music, Mind, Body and Brain, Department of Music, Art and Culture Studies, University of Jyväskylä, Finland.*

[1]geoff.luck@jyu.fi

## Abstract

The author's goal in this paper is to explore how artificial intelligence (AI) has been utilized to inform our understanding of and ability to estimate at scale a critical aspect of musical creativity — musical tempo. The central importance of tempo to musical creativity can be seen in how it is used to express specific emotions (Eerola and Vuoskoski 2013), suggest particular musical styles (Li and Chan 2011), influence perception of expression (Webster and Weir 2005) and mediate the urge to move one's body in time to the music (Burger et al. 2014). Traditional tempo estimation methods typically detect signal periodicities that reflect the underlying rhythmic structure of the music, often using some form of autocorrelation of the amplitude envelope (Lartillot and Toiviainen 2007). Recently, AI-based methods utilizing convolutional or recurrent neural networks (CNNs, RNNs) on spectral representations of the audio signal have enjoyed significant improvements in accuracy (Aarabi and Peeters 2022). Common AI-based techniques include those based on probability (e.g., Bayesian approaches, hidden Markov models (HMM)), classification and statistical learning (e.g., support vector machines (SVM)), and artificial neural networks (ANNs) (e.g., self-organizing maps (SOMs), CNNs, RNNs, deep learning (DL)). The aim here is to provide an overview of some of the more common AI-based tempo estimation algorithms and to shine a light on notable benefits and potential drawbacks of each. Limitations of AI in this field in general are also considered, as is the capacity for such methods to account for idiosyncrasies inherent in tempo perception, i.e., how well AI-based approaches are able to 'think and act like humans.'

## Introduction

Artificial Intelligence (AI) can be defined as the simulation of human intelligence in computer systems programmed to think and act like humans (Russell, Norvig and Crouch 2010). The rapid development of AI and related technologies has had profound effects on a range of everyday and high-level activities and industries. From finance to healthcare, shopping, and transport, AI and its constituent and associated methods have opened a Pandora's box of opportunity across a growing array of societal structures, including commerce, government, and academia. In particular, the impact of AI on creative processes marks perhaps the most significant incursion yet into what we conceptualize as true human activity.

Music is one such creative activity, a universal found in some form in all known human cultures. And recent work has revealed the potential for AI to transform how we understand, evaluate, and model creative musical processes. From composition to performance, recommendation to transcription, AI has been shown to have a range of impacts on this most human of creative endeavors. The author's goal in this paper is to explore how AI has been utilized to inform our understanding of, and our ability to estimate at scale, one particular, fundamental aspect of musical creativity — musical tempo.

## Musical Tempo

Tempo, including its absence, is a fundamental characteristic of any piece of music. Typically indicated in beats per minute (bpm), tempo refers to the speed or pace a musical work is (or is intended to be) played at. The fundamental importance of musical tempo can be seen in how it is utilized by composers and performers to express specific emotions (Eerola and Vuoskoski 2013), to suggest particular musical styles (Li and Chan 2011), and to build and release tension (Goodchild, Gingras and McAdams 2016). From a listeners' perspective, musical tempo can influence perception of expression (Webster and Weir 2005), level of arousal (Lundqvist, Carlsson, Hilmersson, and Juslin 2009), as well as the urge to move one's body in time to the music (Burger et al. 2014).

Musical tempo can be understood as representing two distinct concepts: A *physical* concept referring to the number of events *produced* per minute or a *psychological* concept corresponding to the number of events *perceived* per minute. The latter can be conceptualized as the rate at which a typical listener would tap or move along to a piece of music (Drake, Gros, and Penel 1999; Sachs 1953). The majority of scientific work focuses on perceptual tempo. Indeed, in the field of music information retrieval (MIR), tempo appears to be implicitly understood as the rate at which a listener would tap along to the music (Fraisse 1982). While there exists considerable individual variation in how listeners define the exact tempo of a piece of music (e.g., at which beat level or octave they tap along to), reliable estimation of perceived tempo across large corpora of music remains a key goal in MIR (Böck 2010). This is because tempo plays a crucial role in a range of applied MIR-related pursuits, including music classification (Nieto 2020), genre recognition (Tzanetakis and Cook 2002), emotion analysis (Cambouropoulos 2000), algorithmic generation (Mauch, Durieux, Müller and Riedl 2015), transcription (Boeck and Widmer 2014), sound source separation (Uhlich, Kim and Lee 2017), and recommendation (Zhang et al. 2018). The significance of tempo in the MIR community is further emphasized in the recurring competitions held between tempo extraction algorithms over the past two decades (Downie 2008).

Performance of an algorithm is typically assessed according to standardized metrics such as Accuracy0 (Acc0), Accuracy1 (Acc1), Accuracy2 (Acc2), and, where relevant, ClassAccuracy. Acc0 evaluates the number of (rounded) estimated tempo values that are identical to an annotated ground truth; Acc1 evaluates if estimated tempo lies within

+/- 4% of the annotated ground truth; Acc2 is the same as Acc1 but considers octave errors – confusing the actual tempo to its rhythmic counterparts – as correct; ClassAccuracy evaluates ability to correctly classify tempo from a range of classes where a classification approach is implemented. In terms of frameworks utilized, tempo estimation methods can be roughly divided into traditional approaches and more recent, AI-based methods.

Traditional tempo estimation methods typically detect signal periodicities that reflect the underlying rhythmic structure of the music (Grzywczak and Gwardys 2014), often using some form of autocorrelation of the amplitude envelope (Toiviainen and Lartillot 2007). Examples of autocorrelation-based tempo estimation algorithms include the *Miningsuite* (Lartillot 2019) and *marsyas* (Tzanetakis and Percival 2013). The former enjoys widespread use in areas such as music psychology, the latter in the field of MIR. Both have achieved good results in tempo detection tasks and have been utilized in distinct areas of music information research, including both industry and academia. An alternative to these algorithms is to crawl tempo metadata via Spotify's API. Although Spotify uses a proprietary algorithm not available to the public, it has nonetheless grown to be an industry standard with many uses also in academia. An open-source alternative is *librosa* (McFee et al. 2015), a Python package that can be used to extract temporal and spectral features from audio. Accuracy of these different possibilities varies, and none should be considered the de facto standard.

More recently, AI-based methods utilizing convolutional neural networks (CNNs) or recurrent neural networks (RNNs) on spectral representations of the audio signal have enjoyed significant improvements in accuracy over their traditional counterparts (Aarabi and Peeters 2022). Examples of models employing convolutional and recurrent neural networks include *Schr* (Schreiber and Müller 2018) and *böck* (Böck, Krebs and Widmer 2015), respectively. Other AI-based techniques include those based on probability (e.g., Bayesian approaches, hidden Markov models (HMM)), classification and statistical learning (e.g., support vector machines (SVMs)), and artificial neural networks (ANNs) (e.g., self-organizing maps (SOMs), CNNs, RNNs, DL). This is neither an exhaustive list nor the sole framework within which to categorize these approaches. It does, however incorporate many of the most common methods.

Related to tempo estimation is recognition of rhythm patterns. The latter is really an extension of the former but has received less attention in the literature compared to tempo recognition. This is likely due to the difficulty in creating datasets that are annotated for rhythm patterns, since distinguishing between similar patterns remains a difficult task. As with tempo estimation, early systems for recognizing rhythm patterns used hand-crafted signal processing and statistical models, while more recent systems have used ML and DL techniques. Examples of these include systems that utilise a beat spectrum (Foote 2002), beat histogram (Tzanetakis 2002), harmonic analysis (Peeters 2011), and scale transform (Holzapfel 2011). The latter approach was extended by Marchand and Peeters (2014, 2016) by combining it with the modulation spectrum and adding correlation coefficients between frequency bands.

What follows is an overview of some principal applications of these approaches to modeling and estimating perceived musical tempo and related tasks such as rhythm recognition and beat tracking. The object is to describe in general terms – and examine the level of accuracy achieved by – AI-based tempo estimation algorithms, as well as to shine a light on notable benefits and potential drawbacks of different approaches. Limitations of the use of AI in this field in general are also considered. Attention is paid to how well AI-based methods account for idiosyncrasies inherent in tempo perception, i.e., how well they are able to 'think and act like humans.' Gaps in the literature and future directions for research are also highlighted.

## AI and Musical Tempo

### 1. Reducing Octave Errors

Traditional tempo estimation typically starts with detecting rhythmically related events and then estimating tempo by finding the dominant periodicity of the onsets related to the beat positions (Hainsworth 2004). Various techniques have been used for the latter aspect, including autocorrelation (Percival and Tzanetakis 2014), comb filters (Scheirer 1998), dynamic programming (Ellis 2007), Hidden Markov Models (Klapuri, Eronen and Astola 2006), and source separation (Gkiokas, Katsouros, Carayannis and Stafylakis 2012). Despite sizeable performance increases over the years, most of these approaches still suffer from octave errors. Many attempts have been made to reduce such errors by using techniques such as Gaussian Mixture Models (GMM) (Peeters and Flocon-Cholet 2012), neural networks (Böck, Krebs and Widmer 2015), k-nearest neighbour classification (k-NN) (Wu and Jang 2014), genre classification (Hörschläger, Vogl, Böck and Knees 2015), and Support Vector machines (SVM) (Percival and Tzanetakis 2014). Such methods have often been applied as a separate stage in a multi-stage approach.

For instance, some authors have introduced a machine learning (ML)-driven classification step to help guide the algorithm to the correct tempo octave. Chen, Cremer, Lee, DiMaria, and Wu (2009) took a novel approach to this by using musical descriptors related to perceived mood to train a statistical model of perceived tempo classes in a unique 299-track dataset. Using an SVM and ground truth ratings as input, each track was pre-classified as being either 1) very slow, 2) somewhat slow, 3) somewhat fast, 4) very fast. The logic was that this first-stage classification could then be used to improve the accuracy of *any* conventional tempo estimation algorithm. To illustrate this, they tested the effect of their first-stage octave classifier in combination with a range of existing algorithms on the *ISMIR04* and *MIREX06* datasets. The algorithms selected were the 11 submitted to the ISMIR04 tempo estimation competition plus a new algorithm developed by the authors. Results revealed that the authors' octave-corrected approach indeed led to a significant reduction in octave errors with these algorithms. In fact, performance across all algorithms improved by an average of 45% and 65.5% for the *ISMIR04* and *MIREX06* datasets, respectively. It's worth emphasizing again that there was no additional low-level analysis of temporal events or repetition rates in the audio signal of each track: The corrective procedure was thus entirely independent of tempo-specific information. This represents a major benefit of Chen et al.'s (2009) system. In addition, the fact that it can be used to improve *any* traditional tempo estimation algorithm makes it a very versatile approach. Nonetheless, even with the correction, only three of the algorithms tested achieved Acc1 performance above 50%.

Wu (2015) also tackled the octave error problem by estimating tempo using a two-stage process, albeit in a slightly different fashion. First, the two most dominant tempi were estimated from a tempogram obtained from the short time Fourier transform of an onset detection function (ODF). Second, a k-NN or SVM classifier was used to discriminate the predominant tempo from the pair so identified. Wu (2015) evaluated this approach with the *ISMIR04 Ballroom* and *Songs* datasets. Results revealed a reduced rate of octave errors compared to previous approaches. Specifically, Wu achieved high levels of performance on Acc1 measures for both *Ballroom* (78.5%) and *Songs* (62.6%). In fact, both of these accuracy rates were superior to 7 tested alternatives – *Klapuri* (Gouyon et al 2006; Eronen and Klapuri 2010), *Gkiokas-MA* (Gkiokas, Katsourus and Carayiannis 2012), *Gkiokas* (Gkiokas et al 2012), *Uhle* (Gouyon et al 2006), *Scheirer* (Gouyon et al 2006; Scheirer 1998), *Gainza-Hyb1* (Gainza and Coyle 2011), *Gainza-Hyb2* (Gainza and Coyle 2011). For Acc2, performance was again superior to the same 7 alternatives for *Ballroom* (95%), but not for *Songs* (80.6%). The main benefit of Wu's approach was generally superior performance compared to a range of previous approaches on the two databases selected for testing. The principal drawbacks were that 1) only a small number of datasets were tested, and 2) because of the relative homogeneity of the datasets utilized, it's unclear how the system would perform on more genre-diverse collections.

In a similar vein, Dutta (2018) framed octave error correction as a classification problem, and proposed a four-stage 'tempo estimation plus octave correction' architecture termed *base+octv*. Stage 1 involved detecting onsets; Stage 2 entailed beat period detection; Stage 3 involved Histogram building; and Stage 4 implemented octave classification via an SVM. The SVM utilized 5 features and a non-linear kernel to classify tempo into one of 3 octave classes. Like similar approaches, the goal was to reduce octave errors. Acc1 performance on 6 classic datasets (*Ballroom*, *Hainsworth*, *ACM_MIRUM*, *GTZAN*, *ISMIR04*, *SMC_MIRUM*) was found to be superior to a range of existing approaches (*Percival* (Percival and Tzanetakis 2014), *Klapuri* (Klapuri, Eronen and Astola 2006), *Gkiokas* (Gkiokas et al. 2012), *IBT* (Oliveiraet al. 2010), *qm_vamp* (Davies and Plumbley 2007), *Scheirer* (Scheirer 1998)), while Acc2 was found to be on-par with though not better than the state-of-the-art. Specifically, Acc1 performance was 69.6% (slightly better than the state of the art), while Acc2 performance was 91.2% (on-par with but not better than the state of the art). Dutta argues that their approach might classify unseen audio files better. However, they provide no evidence to support this assertion. The principal drawback is thus that this system was not tested on unseen data.

The approaches outlined above, then, suggest that one way of reducing octave errors is to use classification techniques prior to or after principal tempo estimation to limit the range of possible tempos. The most versatile implementation of this approach seems to be that described by Chen et al (2009) since their classifier can apparently function in combination with any tempo estimation algorithm. Despite classification-based octave error reduction techniques improving performance of (mostly) traditional tempo estimation techniques, the field evolved with the development of neural network-based approaches and the framing of tempo estimation itself as a classification problem.

## 2. Tempo Estimation Itself as a Classification Problem

More recently, the traditional-plus-octave-error-reduction approach has been largely supplanted by steadily-improving architectures based on neural networks. Here, the estimation of tempo itself is framed as a classification problem. Authors have made significant strides utilizing CNNs for this purpose. This work was born out of the use of CNNs in non-MIR fields, such as image classification. The widespread analysis of spectrograms in MIR research, and the visual nature of such time-frequency representations of music, led to standard computer vision CNNs being implemented to 'see' and classify events in spectrograms (e.g., Choi, Fazekas and Sandler 2016; Phan, Hertel, Maass and Mertins 2016; Han, Kim and Lee 2016). It seems noteworthy that most MIR DL scientists used spectrograms as input to their CNNs (Choi et al. 2016; Phan et al. 2016; Han et al. 2016; Pons, Lidy and Serra 2016; Schlüter and Böck 2014). Most even adopted stock rectangular filters for classification. This latter point highlights potential shortcomings of this approach: While images have spatial meaning, spectrograms' dimensions represent time and frequency. Consequently, wider or higher filters might be capable of learning longer temporal dependencies in the audio domain or timbral features spread across a wider range of frequencies, respectively. The result would be *musically*- as opposed to visually-motivated filter shapes. Indeed, this is precisely what encouraged Pons and Serra (2017) to investigate whether MIR CNNs might benefit from a design oriented towards learning musical features rather than 'seeing' spectrograms, i.e., filter shapes adapted to musical concepts.

Pons and Serra (2017) proposed a novel design strategy for convolutional neural networks (CNNs) in music classification tasks, specifically spectrogram analysis. Their approach was to use different filter shapes adapted to fit musical concepts within the first layer. This is more expressive and certainly intuitive than the default small rectangular filters typically used. Pons & Serra (2017) developed a shallow approach in which two complimentary architectures designed to model onsets (*O-net*) or patterns (i.e., rhythm, tempo) (*P-net*) using shorter and longer filters, respectively, were implemented in parallel. By systematically manipulating combinations of the two with different parameter settings, the authors derived 8 different approaches. Performance of these approaches on the *Ballroom* dataset was then compared against 3 DL-based approaches – *Time* (Pons et al. 2016), *Time-freq* (Pons et al. 2016), *Black-box* (Pons et al. 2016) – and 1 non-DL approach – *Marchand et al.* (Marchand and Peeters, 2016). Results showed that Pons & Serra's (2017) strategy was useful for fully exploiting the representational capacity of the first CNN layer when modelling music. Specifically, their best approach scored second only to *Marchand et al.* in terms of tempo estimation accuracy.

Building on Pons and Serra's (2017) work, Schreiber and Müller (2019) further developed the use of CNNs for musical tempo estimation by exploiting the different semantics of spectrograms' time and frequency axes. Train/test datasets utilized were *Ballroom* (Schreiber and Müller 2018), *EBall* (Schreiber and Müller 2018; Gouyon et al. 2006; Marchand and Peeters 2016 b), *GiantSteps Key* (Knees et al. 2015), *GiantSteps Tempo* (Knees et al. 2015; Schreiber and Müller 2018 b), *GTzan Key* (Tzanetakis and Cook 2002; Kraft, Lerch and Zölzer 2013), *GTzan Tempo* (Tzanetakis and Cool 2002; Percival and Tzanetakis 2014), *LMD Key* (Raffel 2016; Schreiber 2017), *LMD Tempo* (Schreiber and Müller 2018; Raffel 2016), and *MTG Tempo/MTG Key* (Schreiber and Müller 2018; Faraldo et al. 2017). Employing a range of approaches, from shallow, domain-specific to deep variants with directional filters, they found that axis-aligned architectures performed on par with VGG-style networks developed for computer vision. At the same time, they were both less affected by confounding factors and required fewer model parameters.

The adoption of neural networks by the MIR community, then, significantly increased the capabilities of tempo estimation algorithms. In particular, approaches that moved beyond 'seeing' spectrograms to applying directional filters based on musical concepts upped the overall level of performance in the field while at the same time simplifying the architectures. Still, these approaches largely only focused on estimating one musical characteristic – tempo – at a time. What came next were approaches able to estimate multiple characteristics – such as tempo, beat, and downbeat – simultaneously.

## 3. Multi-Task Methods for Simultaneous Estimation of Multiple Characteristics

As we've seen so far, early tempo estimation methods based on the application of signal processing techniques such as autocorrelation analysis, comb filtering, and the discrete Fourier transform (DFT), to onset strength signals (OSS) extracted from the audio (e.g., Percival and Tzanetakis 2014; Böck, Krebs and Widmer 2015; Wu, Lee, Jan, Chang, Lu and Wang 2011) suffered from frequent octave confusion. More recently, deep neural networks (DNNs) used for direct tempo estimation, in which the task was framed as a classification problem (Schreiber and Müller 2018; Foroughmand and Peeters 2019) increased overall level of performance. More recently still, multi-task methods have been developed for joint estimation of multiple metrical elements, such as beat and downbeat (Goto 2001; Böck, Krebs and Widmer 2016) and beat, downbeat, and tempo (Böck, Davies and Knees 2019; Böck and Davies 2020). The overlap of the problems within these multi-task approaches have led them to achieve high levels of performance using innovations such as Long Short-Term Memory (LSTM) and Temporal Convolutional Networks (TCNs), as well as exploiting the benefits of training with data augmentation.

TCNs, which first appeared in the WaveNet generative audio model (van den Oord et al. 2016), were proposed by Davies and Böck (2019) as an alternative to a CNN approach using Bidirectional Long Short-Term Memory (BLSTM) for audio-based beat tracking. The authors observed that TCNs achieved state-of-the-art performance on a wide range of existing beat tracking datasets. They were also well suited to parallelization, allowing them to be trained efficiently even on very large datasets. Moreover, they required only a small number of weights. According to the authors, these attributes made TCNs a promising choice for audio-based beat tracking tasks.

Böck, Davies and Knees (2019) built upon Davies and Böck's (2019) beat-tracking system underpinned by temporal convolutional networks (TCNs). The authors proposed a multi-task learning system that simultaneously tracked beats and estimated tempo. As in Davies and Böck (2019) the system used TCNs, but here globally aggregated the skip connections, feeding them into a tempo classification layer. The multi-task nature of the system allowed it to exploit the mutual information of both tasks (by definition, the two tasks are highly interconnected) and improve one by learning *only* from the other. To assess the approach, a range of existing annotated datasets were used for training (*Ballroom*, *Beatles*, *Hainsworth*, *Simac*, *SMC*, *HJDB*) and testing (*ACM Mirum*, *GiantSteps*, *GTZAN*), and performance was evaluated against four reference systems (Gkiokas et al., Percival and Tzanetakis, Böck et al., Schreiber and Müller). Tempo estimation was evaluated with Acc1 and Acc2 measures. For both evaluation criteria, the multi-task approach achieved state-of-the-art performance at (both beat tracking and) tempo estimation. In particular, the system demonstrated improved performance on beat-tracking when trained with data that included tempo-only annotations. In other words, much like humans, the system learnt from information provided concerning a different but related task. However, no mention was made of tempo estimation performance when only beat-tracking data was included.

Böck et al (2019) suggested that this approach may have a significant impact on beat tracking moving forward, as it allows for the use of alternative training data (global tempi) that are more prevalent and easier to annotate. The authors also discussed the computational benefits of the proposed approach, which include efficient training and reduced over-fitting. One potential drawback of Böck et al's approach is that no mention is made of tempo estimation performance when only beat-tracking data is included. This isn't strictly a negative aspect of the system, it's just that no information is provided to clarify this. In terms of being human-like, the authors suggest that further research should be conducted on this "compact" deep model approach for generalization capabilities and re-use of information for end-users on unseen data.

In a conceptual development of Davies and Böck (2019), Oyama, Ishizuka and Yoshii (2021) proposed a phase-aware method for jointly estimating beat *and* downbeat in popular music. They utilised a deep neural network (DNN) that estimated the beat phase at each frame instead of the beat presence probability. Their approach used all frames for training, not just a limited number of beat frames. The authors also modified the post-processing method for the estimated phase sequence. Different multi-task learning architectures for joint beat and downbeat detection were investigated, and the experimental results demonstrated the importance of phase modelling for stable beat and downbeat estimation.

A further multi-task development was proposed by Böck and Davies (2020). They described a state-of-the-art DNN that simultaneously estimated tempo, beat location, *and* downbeat location. In particular, they used a data augmentation approach to expose their network to a wider range of information pertaining to these three aspects. Data augmentation, of course, is a major benefit of AI-oriented approaches in general, allowing a larger and more diverse dataset to be effectively created from a smaller more homogeneous one. The result of Böck and Davies' approach was a performance increase of up to 6% over existing approaches.

The data augmentation argument above notwithstanding, large datasets of tempo-relevant annotations have facilitated development of so-called 'data-driven' tempo estimation in which ML algorithms learn from the annotated data. Initial developments in the field utilized algorithms based on approaches including bags of classifiers (Levy 2011), Gaussian Mixture Models (GMM) (Xiao et al. 2008; Peeters and Flocon-Cholet 2012), k-Nearest-Neighbors (k-NN) (Seyerlehner et al. 2007), Random Forests (Schreiber and Müller 2017), and SVMs (Chen et al. 2009; Gkiokas et al. 2012; Percival and Tzanetakis 2014). More recently, DL approaches have come to dominate the ML space.

An early DL-based tempo estimation system was that proposed by Böck, Krebs and Widmer (2015). This applied a bank of resonating comb filters to the output of an RNN to predict beat position then predict tempo from the estimated periodicity of the signal. Their approach did not use hand-crafted features. Instead, the authors used an intermediate beat-level representation of the signal as input to the comb filter bank. No complex post-processing was applied. Instead, the output was simply the highest resonator's histogram peak. Böck et al's approach achieved state-of-the-art performance on nine out of the ten datasets tested. An alternate method was proposed by Schreiber and Müller (2018) that swapped the comb filter for a mel-spectrogram-based approach again applied to a CNN. The latter approach framed tempo prediction as a classification task into tempo classes.

More recently, Foroughmand and Peeters (2019) developed a hybrid approach combining tempo- and genre-related information into a 'hand-crafted-plus-data-driven' approach they called Deep Rhythm (DR). DR represented a significant development because it simultaneously estimated tempo and classified rhythm patterns/genres. To do this, the authors proposed a new representation of the DNN input called harmonic constant-Q modulation (HCQM) that accounted for tempo frequencies in the harmonic series. DR considers how tempo and rhythm pattern interact to more accurately model the audio input. HCQM represented the harmonic series of tempo candidates in audio signals using a 4D-tensor, which was then used as input for a convolutional network to perform tempo and rhythm pattern estimation. In testing across multiple datasets, Foroughmand and Peeters observed incremental improvements in Acc1 for tempo estimation. At the same time, DR outperformed previous approaches on Acc1 and Acc2 measures, though only for *Ballroom* (ballroom music) and *Giant-steps tempo* (electronic music) test sets. This suggests that, in line with its name, DR performs best when rhythm is more clearly defined. Thus, DR offered incremental improvements in Acc1 for tempo estimation.

Multi-task methods, in which multiple characteristics are estimated simultaneously really demonstrate the power of AI-based approaches in estimating temporal features of music. This is especially true in cases, such as the approach described by Böck et al (2019), in which training on one data type, e.g., tempo annotations, improves performance on a related but different task, e.g., beat tracking. One aspect of tempo estimation not discussed so far, and that has also benefited extensively from AI-driven approaches, is estimation of local vs global tempo.

## 4. Estimation of Local vs. Global Tempo

Most of the literature on tempo estimation, and all that reviewed so far, focuses principally on global tempo, i.e., the average tempo of an entire piece of music. This is likely because a steady, largely isochronous beat is a characteristic of a significant proportion of recorded music. The result is that, in most musical styles, especially of the popular era, tempo remains relatively unchanged throughout a track. This is particularly true of electronic styles such as EDM in which the beat is entirely machine-driven. However, not all music exhibits such tempo(ral) isochrony. More temporally expressive styles, notably those in classical-related genres, can exhibit huge deviations from an 'average' or global tempo. For these styles, estimation of local tempo – the tempo at different moments in time or covering shorter epochs – is critical. Several authors have focused either on estimating local tempo only or joint estimation of both local and global tempo.

Schreiber, Zalkow and Müller (2020), for instance, modelled local tempo in a selection of classical music pieces. As noted above, tempo is known to fluctuate significantly in such music. They found that CNN-based approaches quite accurately captured local tempo even for such expressive classical styles as long as they were trained on the target genre. Importantly, they observed that their results were very dependent on the specific training-test split selected.

In a more sophisticated approach, Schreiber & Müller (2018) trained a CNN to estimate both local *and* global tempo in what they termed a single-step approach. In a traditional setup, note onsets or beats are first identified, from which tempo is then estimated. Schreiber & Müller's approach instead framed tempo estimation as a multi-class classification problem. This permitted the use of a single-step method. Their CNN, trained on a large dataset covering a wide range of genres and tempi, was able to estimate tempo based on less than 12 seconds of audio as input. The ability to estimate tempo on such a relatively short sample of music made their algorithm particularly suitable (with caveats) to estimate not just global but also local tempo.

Schreiber and Müller (2018) compared their approach to the *böck* (Böck, Krebs and Widmer 2015) and *schr* (Schreiber and Müller 2018) algorithms using a standard range of datasets: *ACM Mirum* (Peeters and Flocon-Cholet 2012), *Ballroom* (Gouyon et al 2006), *GTzan* (Tzanetakis and Cook 2002), *Hainsworth* (Hainsworth 2004), *ISMIR04* (Gouyon et al 2006), *GiantSteps Tempo* (Knees et al. 2015), and *SMC* (Holzapfel et al. 2012), the union of which they termed *Combined*. Their *new* approach achieved the highest results in terms of the strict metrics Acc0 (44.8%) and Acc1 (74.2%). For octave-error tolerance (Acc2), *new* (92.1%) was slightly outperformed by both *böck* (93.6%) and *schr* (92.2%), but all approaches performed very well. Results suggested that *new* was better than *böck* at correctly estimating tempo octave, while *böck* and *schr* were better if the metrical level was ignored. In terms of Acc1, *new* was significantly better than both *böck* and *schr* for the *Ballroom* (92.0%), *GiantSteps* (73.0%), and *ACM Mirum* (79.5%) datasets. The finding that *böck* and *schr* outperformed *new* on the more genre-diverse *GTzan* and *Hainsworth* datasets suggested that genre-wide training would improve the latter's results on other datasets as well.

So, Schreiber and Müller's single-step approach compared very favorably to other state-of-the-art techniques in terms of global tempo estimation. However, because it also performed well at local tempo estimation, it was found to be useful for identifying and visualizing tempo drift in musical performances. This latter aspect is particularly useful for music analysis purposes. Further benefits included the fact that it did not rely on handcrafted features, instead being completely data-driven. Perhaps the most notable advantage of the system, though, was how well it dealt with tempo octave confusion. The significant reduction in such errors perhaps demonstrates how well it 'thought and acted like a human.'

There were, nonetheless, a few drawbacks to the system. First, since Jazz, Classical, and Reggae genres were missing from the training data, it remains unclear how it would perform on more genre-diverse datasets. Second, the authors note that the network architecture could be improved by reducing the number of parameters, such as the use of shorter filters, dilated convolutions, residual connections, and a suitable replacement for the fully connected layers. The benefit of these changes, of course, would be to reduce the number of operations needed for training and estimation, making the approach more efficient in the process.

In another multi-method approach, Istvanek (2021) compared conventional and state-of-the-art methods of beat tracking. While most tempo estimation work focuses on music with a broadly isochronous pulse, this study tackled the more complex task of tracking string quartet music. The conventional system tested for this purpose was the beat tracking module in the Python *librosa* library (McFee et al. 2015). Like most earlier approaches, this system tracks periodicity in the onset strength envelope via modification of spectral difference or spectral flux. Such frameworks are known to perform poorly with rapidly changing tempi typical of Western art music, although this problem can be mitigated to some extent using an adaptive window size based not on a fixed number of input values but a fixed number of inter-onset-intervals (Müller, Konz, Scharfstein, Ewert, and Clausen 2009). The state-of-the-art method tested was the madmom Python module.[i] This uses a bidirectional RNN with Long LSTM cells, the latter allowing the network to store information relating to longer-term temporal structure. The idea was that this should permit more accurate detection of beat location. Probability estimation of beat location within frames was accomplished via a dynamic Bayesian network (DBN) approximated by an HMM. In comparing the two approaches, Istvanek (2021) found that while *librosa* was best at estimating average global tempo, the RNN-based madmom module offered a better representation of the rhythmic structure and detected the highest number of individual beats. One question Istvanek (2021) raises is whether neural network-based systems should not be considered conventional today. Given their widespread, even dominant use, this seems like a valid question to ask. Also, musicological analysis requires minimum time spent on manual editing and ground truth annotation. Thus, accurate representation of rhythmic structure, especially number and position of beats, is crucial in any beat tracking system for this type of analysis.

With increasing complexity and power, AI approaches have brought us a long way in tempo estimation in little more than a decade. In particular, methods of simultaneous estimation of both multiple characteristics and tempo across multiple time frames (local, global) have proved extremely successful. Nonetheless, architectures continue to increase in complexity, and the current state-of-the-art can provide tempo estimations bordering on perfect. In the final section, three particular state-of-the-art approaches are surveyed to give a flavor of where AI-driven tempo estimation stands today.

## 5. Increasing Complexity and State-of-the-Art

de Souza, Moura and Briot (2021) compared tempo estimation performance of two systems based on artificial neural networks — an architecture utilizing a Bidirectional Recurrent Neural Network (B-RNN) that takes as its input the Mel spectrogram and which outputs the estimated tempo class, and Schreiber and Müller's (2018) Convolutional Neural Network (CNN) architecture. Both networks were trained on the same extensive database (12,550 tracks), including percussion-only tracks, and compared both with each other and with Schreiber and Müller's original(ly trained) algorithm. Results revealed that the B-RNN model performed on-par with (though in most cases did not outperform) the CNN-based approaches, but was particularly accurate for percussion-only tracks. This suggests that rhythmic elements play a mediating factor in the ability of neural networks to learn and predict musical tempo. As the authors note, further research on this is hampered by a lack of percussion-focused databases for testing and training.

Song and Wang (2022) paired a hidden-Markov model (HMM) with a periodic recurrent neural network (PRNN) in an attempt to reduce computational complexity in a beat-tracking task. A significant such reduction was achieved by exploiting the frequency contents of the music signal via Fourier transform. Compared to previous implementations of artificial networks such as bidirectional recurrent neural networks (Bi-RNN) and temporal neural networks (TCN), neither of which can perceive the frequency of the musical beat, the HMM-with-PRNN approach achieved close to state-of-the-art performance but with significantly lower computational cost. Indeed, in additional to the high level of performance, this lower computational cost with state-of-the-art performance is clearly the main benefit of Song and Wang's (2022) approach.

As discussed above, Foroughmand and Peeters (2019) developed a hybrid approach combining tempo- and rhythm pattern-related information into a 'hand-crafted-plus-data-driven' approach they called Deep Rhythm (DR). Recently, Aarabi and Peeters (2022) extended Deep Rhythm to simultaneously estimate both tempo and genre of a musical signal. This is possible because tempo and genre are highly correlated aspects of music. The system used a harmonic representation of rhythm as input to a CNN, processed through complex-valued convolutions to consider relationships between frequency bands. A multitask learning approach was then used to jointly estimate tempo and genre. Additionally, a second input branch was added to the system, using a mel-spectrogram input dedicated to timbre, which improved the performance of both tempo and genre estimation.

The authors' network was trained on 8596 tracks from 3 datasets (*Extended Ballroom (EBR)* (Marchand and Peeters 2016), *tempo MTG* (tMTG) (Faraldo et al. 2017), *tempo LMD* (tLMD) (Raffel 2016)) and tested on a total of 3611 tracks across 7 independent datasets (*ACM* (Peeters and Flocon-Cholet 2012), *ISMIR04* (Gouyon et al. 2006), *Ballroom (BR)* (Gouyon et al. 2006), *Hainsworth (Hains)* (Hainsworth 2004), *GTzan* (Marchand et al. 2015), *SMC* (Holzapfel et al. 2012), *Giantsteps (GST)* (Knees et al. 2015) as well as their union (*Combined*). Global tempo accuracy and octave errors were evaluated on Acc1 and Acc2 measures, respectively. Performance of several versions of the extended approach was compared against the original Deep Rhythm. A performance increase was observed cross all test datasets, with Acc1 of up to 97.7% and Acc2 of up to 99.4% (both on the *GST dataset*). Mean improvement across all datasets was 11.2% and 7% for Acc1 and Acc2, respectively. The biggest improvements were apparent for the *SMC dataset* (16.1% and 23.1% for Acc1 and Acc2, respectively), although performance on this dataset was still sub-par compared to all others. The chief benefits of Aarabi and Peeters' (2022) DR approach was simultaneous estimation of both tempo and genre of music. While these two aspects are related, it can be seen that AI-based approaches are being developed which can capture increasingly disparate musical features. The overall improvement of several percentage points on most datasets was another clear win for DR. The only real drawback was the sub-par performance on the *SMC* dataset.

## Summary and Conclusions

Instilling in machines the ability to 'think and act like humans' is a defining objective of AI. The author's goal in this paper has been to survey how AI-based approaches have been utilized to inform understanding of and ability to estimate at scale perception of a fundamental aspect of music, namely its underlying tempo. Given individual variation in how humans define the exact tempo of a piece of music (e.g., at which beat level or octave they tap along to), one way of demonstrating human-like behavior might for a machine to mimic these idiosyncrasies in a convincing manner. Increasing temporal accuracy (Acc0 and Acc1 evaluation) and reducing octave errors (Acc2 evaluation) might thus be considered hallmarks of such behavior. Other indicators might include reduced complexity and increased efficiency. A range of tempo estimation system have been surveyed here that have progressively dealt with these idiosyncrasies and design characteristics. In the process, these algorithms have come to exhibit human-like qualities to an increasingly sophisticated degree.

Building on earlier signal processing-oriented methods of tempo estimation based on detection of periodicities in the music, a range of AI-related techniques have been experimented with, including those based on probability (e.g., Bayesian approaches, hidden Markov models (HMM)), classification and statistical learning (e.g., support vector machines (SVM)), and artificial neural networks (ANNs) (e.g., self-organizing maps (SOMs), CNNs, RNNs, deep learning (DL)). Along the way, specific variants of some of these approaches have been developed. These continually improving approaches have, one might argue, steadily encroached upon what is widely regarded a unique human capacity. So how much more accurate can tempo estimation get?

In 2019, Schreiber, Urbano and Müller posed the question of whether we're done yet with tempo estimation. Given the recent, near-perfect results achieved by the likes of Aarabi and Peeters (2022), for instance, which even then leave some small room for improvement, one might argue that the answer to that question is 'No'. Will we ever be? In light of the myriad peculiarities of the human condition, and despite the attraction and unarguable utility of flawless imitation of human intelligence by a machine, even an 'intelligent' one, perfect estimation of perceptual tempo on a genuinely diverse range of music seems likely to remain forever just beyond reach.

[i] https://pypi.org/project/madmom/